\documentclass[article]{aa}
\usepackage{natbib}
\usepackage{graphicx}
\usepackage{color}
\begin{document}
   \title{Probing quiet Sun magnetism\\
          using MURaM simulations and Hinode/SP results: support for a local dynamo}
   \titlerunning{The quiet Sun magnetism: simulations vs. observations}
   \author{S. Danilovic \inst{1} \and M. Sch{\"u}ssler \inst{1} \and
           S.K. Solanki \inst{1,2}}

   \institute{Max-Planck-Institut f\"ur Sonnensystemforschung,
              Max-Planck-Stra{\ss}e 2, 37191 Katlenburg-Lindau,
              Germany \and
              School of Space Research, Kyung Hee University,
              Yougin, Gyeonggi 446-701, Korea}

   \date{\today}

  \abstract
  {Owing to the limited spatial resolution and the weak polarization signal coming from the quietest regions on the Sun,
  the organization of the magnetic field on the smallest scales is largely unknown.}
  {We obtain information about the magnetic flux present in the quiet Sun by comparing radiative MHD simulations with observations,
  with particular emphasis on the role of surface dynamo action.}
  {We synthesize Stokes profiles on the basis of the MHD simulation results.
  The profiles are degraded taking into account the properties of the spectropolarimeter (SP)
  on board of the Hinode satellite. We use simulation runs with different
  magnetic Reynolds numbers (R$_{m}$) and observations at different heliocentric angles with
  different levels of noise.}
  {Simulations with an imposed mixed-polarity field and R$_{m}$ below the
  threshold for dynamo action reproduce the observed
  vertical flux density, but do not display a sufficiently high horizontal
flux density. Surface dynamo simulations at the highest R$_{m}$
feasible at the moment yield a ratio of the horizontal and
vertical flux density consistent with observational results, but
the overall amplitudes are too low. Based on the properties of the
local dynamo simulations, a tentative scaling of the magnetic
field strength by a factor $2-3$ reproduces the signal observed in
the internetwork regions. }
  { We find an agreement with observations at different heliocentric angles. The mean field strength in internetwork,
  implied by our analysis, is roughly $170$~G at the
  optical depth unity. Our study  shows that surface dynamo could be responsible for most of the
  magnetic flux in the quiet Sun outside the network given that the extrapolation to higher R$_{m}$ is valid.  }
\keywords{Sun: granulation, Sun: photosphere}

\maketitle
%

\section{Introduction}

The origin of the small scale magnetic flux found in the quiet Sun
is uncertain \citep{deWijn:etal:2008}. The estimated order of
magnitude for the magnetic Reynolds number of the granulation flow
indicates that a substantial part of the magnetic field in the
quiet Sun could be generated locally through dynamo action driven
by near-surface convective flows
\citep{Petrovay:Szakaly,Cattaneo:1999,Voegler:Schuessler:2007}.
The simulations suggest that the magnetic field could be organized
in mixed-polarity structures down to very small spatial scales.
The simulations exhibit a mostly horizontal field in the
photospheric layers in the form of low-lying loops connecting
nearby opposite polarities \citep{Schuessler:Voegler:2008}.

Substantial observational evidence has been gathered that
internetwork flux is dominated by strongly inclined, almost
horizontal magnetic fields
\citep{Orozco:etal:2007a,Orozco:etal:2007b,Lites:etal:2008}. The
validity of the deduced properties has been questioned by
\citet{Asensio:2009} who argues that the influence of noise has
not been adequately taken into account in such studies.
Nevertheless, several authors report loop-like horizontal field
structures of different sizes
\citep{Martin:1988,Marian:etal:2007,Harvey:etal:2007,Centeno:etal:2007,Ishikawa:etal:2008}.
Such structures could be due to local dynamo action, but
small-scale flux emergence \citep{Cheung:etal:2008} or flux
expulsion of a pre-existing field by granular flows
\citep{Steiner:etal:2008} probably also contribute to the
horizontal flux.\begin{figure*}
    \centering
    \includegraphics[width=0.85\hsize,angle=90]{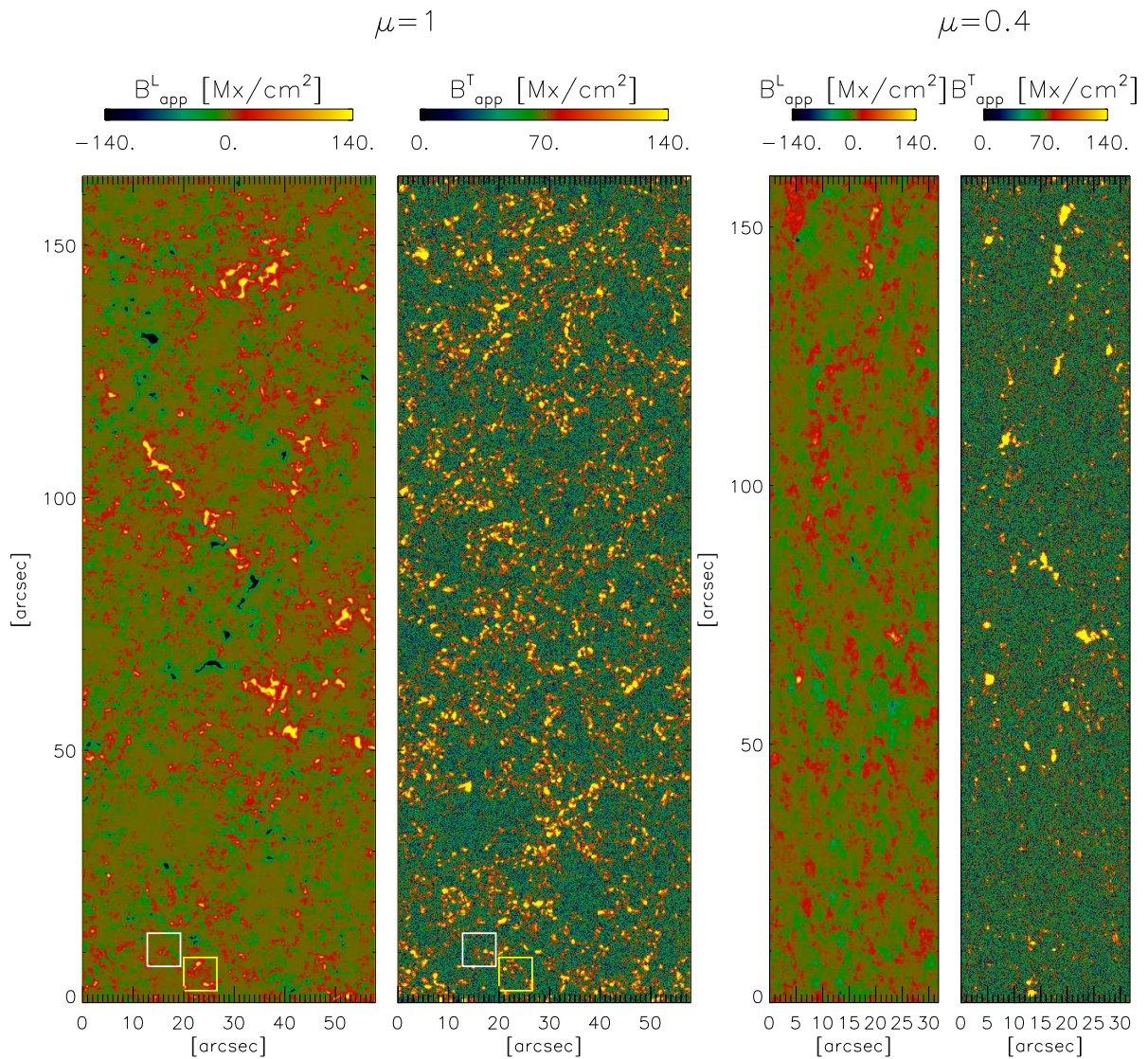}
    \caption{Maps of longitudinal ($B_{app}^{L}$) and transversal ($B_{app}^{T}$) apparent flux density for the observed data sets I (left) and III (right) from Table~\ref{tab:obs}. The small squares indicate
    the sizes of the MHD snapshots. They also mark the position of the regions to which synthesized maps from simulation data are compared (Figures~\ref{fig:maps_obs_285} and ~\ref{fig:maps_obs_702}).}
    \label{fig:maps}
\end{figure*}

Another open question concerns the amount of magnetic flux
contained in the internetwork. The observations give a wide range
of values that vary with the spatial resolution and the diagnostic
technique used (e.g,
\citet{DC:etal:2003,Khomenko:etal:2003,Berdyugina:Fluri:2004,Asensio:etal:2007}).
From many results obtained \citep[see][for
overviews]{deWijn:etal:2008,Solanki:2009}, we mention here only
those that directly concern the work presented in this paper.
Comparing the center-to-limb observations of the scattering
polarization in the Sr I $4607$~\AA~line with the signal
synthesized from the 3D hydrodynamical simulations,
\citet{Trujillo:etal:2004} inferred that the mean strength of the
internetwork field is $\langle B\rangle\sim100$~G, under the
assumption that the magnetic field is isotropically tangled at
subresolution scales and that it fills the whole resolution
element. \citet{Khomenko:etal:2005a}, on the other hand, compared
the observed Stokes $V$ amplitudes of the visible and infrared Fe
I lines with the profiles synthesized from 3D radiative MHD
simulations, and concluded that $\langle B\rangle\sim20$~G (see
also \citet{Nazaret:etal:2009}). \citet{SA:etal:2003} used
Boussinesq 3D simulations of local dynamo action to suggest that a
snapshot with mean longitudinal magnetic field of $50$~G can
reproduce both the observed Hanle and Zeeman signals. However, the
simulations used by these authors are rather idealized and the
solar atmospheric structure was arbitrarily introduced for
comparison with the observational data.

In this paper, we use 3D radiative MHD simulations of the solar
photosphere to obtain an estimate of the true magnetic flux
density in the quiet Sun. Our approach differs from that of
\citet{Khomenko:etal:2005a} in that we also consider fields
produced by a local dynamo action. Also, the synthesized
polarization signals are compared with the higher resolution data,
obtained with the spectropolarimeter \citep{Lites:etal:2001} of
the Solar Optical telescope \citep{Tsuneta:etal:2008} on board of
the Hinode satellite \citep{Kosugi:etal:2007}. The comparison with
the simulations at the highest R$_{m}$ feasible at the moment also
allows us to gain insight into the role of the local dynamo action
for the quiet Sun magnetism.


\section{Observations}

\begin{table*}
\caption{Details of the three Hinode/SP data sets used. The last
three columns, from left to right, give the mean values of the
transversal (with respect to the line of sight), unsigned
longitudinal and signed longitudinal magnetic flux density,
respectively.} \label{tab:obs} \centering
\begin{tabular}{l c c c c c c c c c c}
\hline \hline
 & date & time & FOCUS & $\mu$ & size & exposure & noise & $\langle B^{T}_{app}\rangle$ & $\langle |B^{L}_{app}|\rangle$ & $\langle B^{L}_{app}\rangle$  \\
 & [dd/mm/yy]& UT& & & [arcsec] & time [s] & $10^{-3} I_{c}$ & Mx/cm$^2$ & Mx/cm$^2$ & Mx/cm$^2$\\
\hline
set I & $10/09/07$ & 08:00:00 & 2031 & 1 & $58 \times 164$ & 9.6 & 0.8 & 56.7 & 9.7 & 1.6 \\
set II & $27/02/07$ & 00:20:09 & 2048 & 1 & $0.16 \times 164$ & 67.2 & 0.25 & 54.6 & 11.0 & 1.7 \\
set III & $09/09/07$ & 13:05:05 & 2029 & 0.4 & $160 \times 32$ & 9.6 & 1 & 53.5 & 6.8 & 0.3 \\
\hline
\end{tabular}
\end{table*}

We consider three data sets obtained with the spectropolarimeter
(SP) on board Hinode. Details are given in Table~\ref{tab:obs}.
Data sets I and II were recorded near disc center, while data set
III was obtained closer to the limb. Data sets I and III were
obtained in the scan mode of the Hinode/SP, with an exposure time
of $9.6$~s per slit position. They cover quiet Sun regions at the
disk center and a region near the south solar pole, respectively.
In the case of data set III, we use a $32\arcsec$ wide strip
perpendicular to the scan direction that corresponds to $\mu=0.4$
($\mu$ being the cosine of the heliocentric angle). Corrections
for various instrumental effects are made using the procedure
sp\_prep, included in the
SolarSoft\footnote[1]{http://www.lmsal.com/solarsoft/} package.
The procedure gives the longitudinal and transversal magnetic flux
density maps \citep{Lites:etal:2008} shown in Fig~\ref{fig:maps}.
The mean values are given in Table~\ref{tab:obs}. The rms
continuum contrast values for data sets I and III are $7.5$\% and
$5.1$\% respectively.

Data set II is a time series recorded with fixed slit position. It
consists of $103$ scans at solar disk center, each with an
exposure time of $9.6$~s. After applying a temporal running mean,
the effective exposure time becomes $67.2$~s, which gives a
significantly lower level of noise. This data set has previously
been used by \citet{Lites:etal:2008} and
\citet{Orozco:etal:2007b}.

\section{Simulation data}

We use results from 3D radiative MHD simulations of a layer
containing the solar surface, carried out with the MURAM code
\citep{Voegler:2003, Voegler:etal:2005}. Non-grey LTE radiative
transfer and partial ionization are taken into account. We compare
snapshots from several simulation runs. The basic properties of
the computational domains are given in Table~\ref{tab:domains}. In
all runs, the top of the simulation box is located about 500 km
above the average height level of optical depth unity. The side
boundaries are periodic, whereas the bottom boundary is open,
permitting free in and outflow of matter. The magnetic field is
vertical at the top and bottom boundaries.

\begin{table}
\caption{The simulation parameters. Given are the size of
computational domain and vertical ($\delta{z}$) and horizontal
($\delta{x}$) grid spacing.} \label{tab:domains} \centering
\begin{tabular}{l c c c}
\hline \hline
Run & height/width  & $\delta{z}/\delta{x}$ & R$_{m}$ \\
&  [Mm] & [km] & \\
\hline
mixed polarity & 1.4/6.0 & 14/20.8 & $\sim300$\\
dynamo C& 1.4/4.86 & 10/7.5 & $\sim2600$\\
dynamo G & 1.4/4.86 & 7/5 & $\sim5200$\\
\hline
\end{tabular}
\end{table}

The first run, henceforth referred to as the 'mixed polarity' run,
simulates the decay of the magnetic field in a mixed polarity
region. In this run, local dynamo action
\citep{Voegler:Schuessler:2007} does not occur since the magnetic
Reynolds number is below the threshold for dynamo action. The run
starts with a vertical magnetic field of $|B| = 200$~G, in a
checkerboard-like $2\times2$ pattern, with opposite polarities in
adjacent parts. The field is concentrated and redistributed by the
convective motions; the opposite polarities are pushed together,
which results in flux cancellation and almost exponential decrease
of the mean magnetic field strength. The snapshots with
$\langle|B|\rangle = 35$~G and $\langle|B|\rangle = 20$~G averaged
over the surface $\tau=0.1$ are used in our study. The same
snapshots were used by
\citet{Khomenko:etal:2005a,Khomenko:etal:2005b}.

\begin{figure}
    \centering
    \includegraphics[width=0.7\hsize,angle=90]{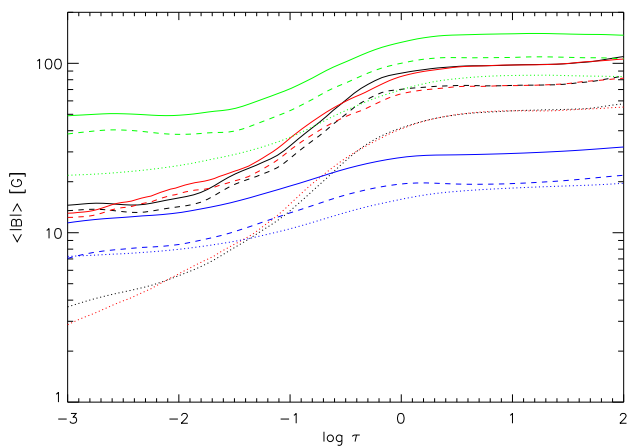}
    \caption{Average total (solid), horizontal (dashed) and vertical (dotted) mean magnetic field strength from MURaM simulations as a function of the optical depth.
    Snapshots from different runs are indicated by line color:
    mixed polarity (blue), dynamo G (red), dynamo C multiplied by factor of $1.5$ (black) and dynamo C with background unipolar field (green).}
    \label{fig:b_comp}
\end{figure}

The second and third group of snapshots are taken from the runs
with a magnetic Reynolds number of the flow sufficiently high for
small-scale turbulent dynamo action to take place. A seed field of
$|B_{0}|=10$~mG grows exponentially in time until the saturation
level is reached. In the dynamo run C, described in
\citet{Voegler:Schuessler:2007}, the mean magnetic field strength
at this phase is $\langle|B|\rangle =23$~G at $\tau=0.1$. In the
case of the dynamo run G, the higher magnetic Reynolds number
leads to a saturation level of the magnetic energy which is a
factor of $1.7$ higher than in run C. Figure~\ref{fig:b_comp}
shows the mean magnetic field strength (averaged over surfaces of
constant $\tau$) as a function of the optical depth for one
snapshot from each run. The field from dynamo C run is multiplied
by a factor 1.5, in order to illustrate the similarity of the
scaled optical depth profiles in the dynamo runs. This factor
differs only by $\sim 15$\% from $\sqrt{1.7}$, the square root of
the corresponding total magnetic energy ratio. This indicates that
a tentative scaling of the dynamo-generated field to represent
higher Reynolds numbers is not completely unreasonable. The dashed
lines show the optical depth profiles of the
 horizontal magnetic field, i.e.,
$\langle B_{hor}\rangle=\langle \sqrt{B_{x}^{2} +
B_{y}^{2}}\rangle$ and the dotted lines represent the average
vertical field $\langle |B_{vert}|\rangle$. The dynamo runs show a
significantly larger $\langle B_{hor}\rangle$ than $\langle
|B_{vert}|\rangle$. Their ratio reaches values between 2 and 4 in
the optical depth interval $-2<\log \tau<-1$. The mixed polarity
snapshot, on the other hand, has a $\langle B_{hor}\rangle$ and
$\langle |B_{vert}|\rangle$ of similar magnitude over all heights.

The last group of snapshots is taken from a simulation run that
continues dynamo run C with a superposed unipolar vertical field
(green lines in Figure~\ref{fig:b_comp}). Such a superposition
might be a way of describing a weak network region. The mean
vertical field strength is thereby increased to around $36$~G at
$\tau=0.1$ and the strength of the horizontal field to around
$50$~G at $\tau=0.1$.

\begin{figure*}
    \centering
    \includegraphics[width=0.3\hsize,angle=90]{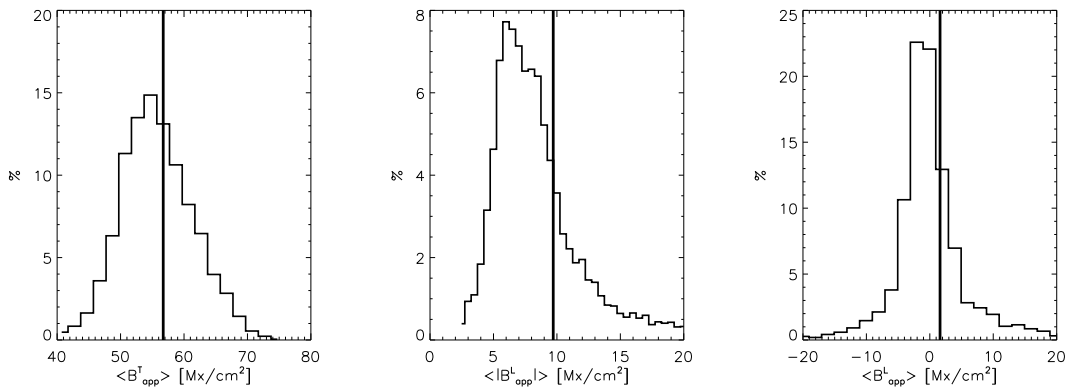}
    \caption{Histograms of the mean horizontal (left), the unsigned vertical (middle) and the signed vertical (right)
    magnetic flux density calculated for $7\arcsec\times7\arcsec$ size regions covered by dataset I.
    The vertical lines mark the mean values for the whole field of view. }
    \label{fig:hist}
\end{figure*}

\section{Spectral synthesis}

The simulation results have been used as input for the LTE
radiative transfer code SPINOR \citep{Frutiger:etal:2000} to
synthesize the Stokes profiles for the heliocentric angles~
$\theta=0^{\circ}$ and $66^{\circ}$ (corresponding to $\mu=1$ and
$\mu=0.4$, respectively). The spectral range that contains the
Fe~I lines at $630.15$ and $630.25$~nm is sampled in steps of
$7.5$~m\AA. The Fe abundance used in the synthesis has been taken
from \citet{Thevenin:1989} and the values of the oscillator
strengths from the VALD database \citep{Piskunov:etal:1995}. We
then applied a realistic point spread function (PSF,
\citet{Danilovic:etal:2008}) to the maps of synthesized Stokes
profiles. The PSF takes into account the basic optical properties
of the Hinode SOT/SP system and a small defocus. Applying an ideal
PSF without defocus reduces the original rms contrast values of
the simulated continuum map from $14.5$\% to $8.5$\%
($\theta=0^{\circ}$) and $11.2$\% to $5.5$\%
($\theta=66^{\circ}$), respectively. For the comparison with the
observations at $\theta=0^{\circ}$, we used a defocus of
$-1.5$~mm, which degrades the continuum contrast of the simulation
to the observed value of $7.5$\%. A value of $-0.75$~mm is used
for the synthesized data at $\theta=66^{\circ}$ in order to match
the continuum contrast of $5.1$\% deduced from dataset III. The
difference in the focus of the SOT between data sets I and III
amounts to approximately 2 steps of the SOT focus mechanism.
Taking into account that there is an uncertainty of one focus step
in the focusing mechanism, the amounts of defocus obtained from
reproducing the observed data sets seem plausible.

The appropriate PSF is applied to the 2D maps at each wavelength
position for every Stokes parameter. The degraded maps are then
rebinned to the pixel size of the Hinode/SP detector. To take into
account the spectral resolution of the spectropolarimeter, the
profiles are convolved with a Gaussian function of $25$~m\AA~FWHM
and resampled to a wavelength spacing of $21.5$~m\AA. Next, a
noise level corresponding to the observations is added, and
finally the procedure by \citet{Lites:etal:2008} is used to
calculate the longitudinal and transversal apparent magnetic flux
densities, $B^{L}_{app}$ and $B^{T}_{app}$.

\section{Results}

We present our results in the following sequence: (1) snapshots
from the simulation runs are compared with the disc center
observations in terms of the mean values of $B^{L}_{app}$ and
$B^{T}_{app}$; (2) the spatial distribution of the magnetic
features in the observed and in the synthesized maps is compared;
(3) a comparison of the probability density functions (PDFs) from
the simulations and from the low-noise data set II is presented.
Finally, we show how the mean magnetic flux density changes with
the heliocentric angle and compare the synthesized and observed
signal at a heliocentric angle of $\theta=66^{\circ}$.

\subsection{Comparison with the observations at disc center}
\label{sec:dc}

\begin{table*}
\caption{Mean apparent magnetic flux densities from observations
and simulations. The values in parentheses are the mean values at
the original resolution of the simulations, before spatial
smearing and addition of noise.} \label{tab:means} \centering
\begin{tabular}{l c c c c c}
\hline \hline
Run & $\langle B^{T}_{app}\rangle$ & $\langle |B^{L}_{app}|\rangle$ & $\langle B^{L}_{app}\rangle$ & $\langle B^{T}_{app}\rangle/\langle |B^{L}_{app}|\rangle$ \\
\hline
observations & 55 & 6 & -2 & 9.2\\

mixed polarity ($30$~G)& 39 (24) & 13 (19) & $-0.2$ ($-0.3$) & 3.0 (1.2)\\
mixed polarity ($20$~G)& 37 (15) & 6.5 (9.0) & $-0.6$ ($-0.3$) & 5.7 (1.8)\\
dynamo C & 36-37 (21-28) & 2.7-3.0 (6.4-7.8) & $-0.1-0.0$ ($-0.2-0.0$) & 12-13 (3.3-3.6)\\
dynamo C (+uni. field) & 45-51 (51-65) & 9-15 (18-27) & 9-14 (9-16) & 3.4-5 (2.4-2.8)\\
dynamo G & 39 (39) & 3.0 (9.6) & 0.1 (0.1) & 13 (4.1)\\\hline
noise & 36 & 2.3 & 0.0 & \\\hline\hline
dynamo C ($mf=3$) & 51-60 (70-81) & 5.3-6.0 (19-24) & $-0.1-0.0$ ($-0.1-0.0$) & 9.6-10 (3.4-3.7)\\
dynamo G ($mf=2$) & 52 (77) & 4.7 (19) & 0.2 (0.2) & 11 (4.0)\\
\hline
\end{tabular}
\end{table*}

When comparing the observational signals with the ones synthesized
from the simulations, one has to take into account that the solar
surface area covered by the observations of data set I is much
larger than that comprised by the simulations. The small squares
in Fig~\ref{fig:maps} indicate the actual size of the simulation
snapshots. In order to take into account this proportion, we
divide the region covered by the observations into subdomains of
$7\arcsec\times7\arcsec$ size. For each of these we calculate  the
mean transversal $\langle B^{T}_{app}\rangle$, the mean unsigned
longitudinal $\langle |B^{L}_{app}|\rangle$ and the mean signed
longitudinal $\langle B^{L}_{app}\rangle$ apparent magnetic flux
densities. The histograms of these quantities are shown in
Fig~\ref{fig:hist}. By considering the contribution of each
subdomain separately, we can exclude the contribution of the
network, which extends the wings of the histograms towards higher
values of $\langle B^{T}_{app}\rangle$ and $\langle
|B^{L}_{app}|\rangle$. The maxima of the distributions, on the
other hand, give an estimate of the typical magnetic flux density
value in the internetwork regions. Consequently, when the whole
observed region is taken into account, the mean values (marked by
the vertical lines in Fig~\ref{fig:hist}) are higher than the
values retrieved from the maxima. The mean values over the whole
FOV are in agreement with the values obtained by
\citet{Lites:etal:2008} and show a ratio of $\langle
B^{T}_{app}\rangle/\langle |B^{L}_{app}|\rangle=5.8$.

The values retrieved from the distribution maxima are given in the
first row of Table~\ref{tab:means}. The other numbers are the
results obtained from the synthesized Stokes profiles from the
simulation snapshots; values determined from the snapshots at
their original resolution are given in parenthesis. They roughly
correspond to the mean vertical and horizontal field strengths
near $\tau=0.1$ (cf. Figure~\ref{fig:b_comp}). Thus, the ratio
$\langle B^{T}_{app}\rangle/\langle B^{L}_{app}\rangle$ obtained
from the simulations at the original resolution reflects the ratio
of the underlying magnetic fields at  $\tau=0.1$. The values are
close to unity for the mixed-polarity run and about $3$ for the
dynamo snapshots. For the dynamo snapshots, the values $\langle
|B^{L}_{app}|\rangle$ after spatial smearing and application of
noise are considerably reduced compared to the noise-free,
unsmeared case owing to the presence of mixed polarities on very
small scales. On the other hand, $\langle B^{T}_{app}\rangle$ is
increased for all groups of snapshots, except for the dynamo run C
with unipolar background field. This is the result of the noise
which is added to the Stokes profiles in order to simulate the
Hinode/SP observations. The row labeled "noise" in
Table~\ref{tab:means} gives values determined from pure white
noise with a standard deviation corresponding to the noise level
of data set I. We give the mean values from 100 realizations. The
value of $\langle B^{T}_{app}\rangle$ determined from pure noise
is almost as large as the values retrieved from the dynamo
snapshots, with the exception of the dynamo run C with a unipolar
background field. This means that hardly any signal of
$B^{T}_{app}$ remained above the noise after spatial smearing and
introduction of noise. Dynamo C run with a unipolar field is a
special case because it has a much higher mean field (cf.
Figure~\ref{fig:b_comp}). The more magnetic flux is introduced,
the more field can be tangled by the turbulent flows, hence
noticeably more horizontal field is generated.

The last two rows of Table~\ref{tab:means} show the results of the
attempt to estimate how much field the dynamo simulations would
have to contain to reproduce the observed $\langle
B^{T}_{app}\rangle$ and $\langle |B^{L}_{app}| \rangle$ values.
Multiplying the magnetic field strengths by factors of $2$ and $3$
(everywhere in the simulation box) in the case of the dynamo run G
and dynamo run C, respectively, gives a mean total magnetic field
strength of $170$~G and $67$~G at the levels of $\tau=1$ and
$\tau=0.1$, respectively. The strength of the mean vertical
magnetic field at the same levels becomes $84$~G and $27$~G,
respectively. These values are consistent with the extrapolation
by \citet{Graham:etal:2009}on the basis of Hinode data. Also, as
we have seen in Fig~\ref{fig:b_comp}, the average mean field
strength distribution of run G can be reproduced by multiplying
the run C result by a factor roughly corresponding to the square
root of the ratio of the total magnetic energies. This suggests
that such a simple scaling might extend somewhat into the R$_{m}$
regime that is not covered by the simulations. The necessary
scaling factor of 2 for run G is not unreasonably large.

\begin{figure}
    \centering
    \includegraphics[width=0.98\hsize,angle=90]{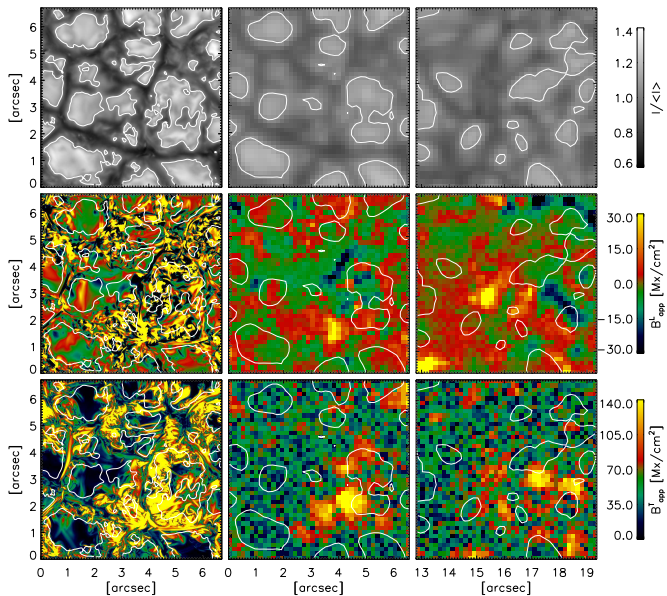}
    \caption{Comparison of a snapshot from dynamo run C  with magnetic field scaled by a factor 3 at original (left column) and Hinode (middle column) resolution, with Hinode observations (right column).
    \textit{From top to bottom}: normalized intensity, longitudinal and transversal apparent magnetic flux density.
     The observed region is outlined by a white square in Fig~\ref{fig:maps}. White lines outline the contours of normalized continuum intensity equal to $1.05$.}
    \label{fig:maps_obs_285}
\end{figure}

Figure~\ref{fig:maps_obs_285} shows one snapshot from the dynamo
run C with the field multiplied by a factor of 3, before and after
spatial smearing, together with an observed region of the same
size, outlined by the white square in Fig~\ref{fig:maps}. Maps of
the longitudinal $B^{L}_{app}$ and transversal $B^{T}_{app}$
magnetic flux density are shown. The granular pattern is indicated
by white contours of normalized continuum intensity equal to
$1.05$. Vertical fields with mixed polarities on scales of less
than a third of an arcsec are smeared into unipolar 'tube' or
'sheet'-shaped patches (yellow and blue in the central frame). The
bundles of horizontal field, composed of loops of different sizes,
are molded into patches with higher $B_{app}^{T}$ values. These
can occur on the edges of granules or between patches with
vertical field of opposite polarity (see, for instance, the
feature at $[4.5\arcsec,2\arcsec]$). Similar features can be seen
in the Hinode maps, e.g. in the two lower frames on the right (see
also \citet{Ishikawa:etal:2008} or \citet{Lites:etal:2008}). At
the position [$5\arcsec,3.5\arcsec$] in the central frame, a
vertical field of $B_{app}^{L}\approx20$~Mx/cm$^{2}$ is located
inside a granule. A similar case has been observed by
\citet{Orozco:etal:2008:vfig}.

\begin{figure}
    \centering
    \includegraphics[width=0.98\hsize,angle=90]{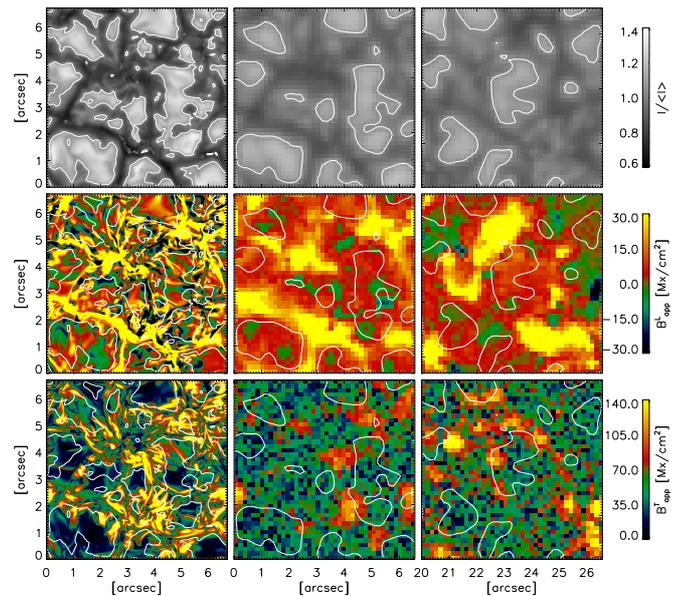}
    \caption{Same as Fig.~\ref{fig:maps_obs_285} but for the dynamo run C with a background unipolar field and a different region from the observed map
    (outlined by a yellow square in Fig~\ref{fig:maps}).}
    \label{fig:maps_obs_702}
\end{figure}

Figure~\ref{fig:maps_obs_702} shows a different region on the Sun,
with more vertical flux (outlined by the yellow square in
Fig~\ref{fig:maps}) compared with a snapshot from the dynamo run C
with a unipolar background field. No scaling of the dynamo field
has been carried out here. The maps based on the simulation show
features which are very similar to the structures present in the
observed weak network region, although the $B_{app}^{T}$ signal is
somewhat weaker than in the observations. Small patches of
horizontal field correspond to small loops that are visible
between concentrations of the vertical flux at original resolution
of the simulations.

\subsection{Comparison with the data set II}
\label{sec:setII}

\begin{figure}
    \centering
    \includegraphics[angle=90,width=0.98\hsize,trim= 0cm 0cm 0cm 0cm,clip=true]{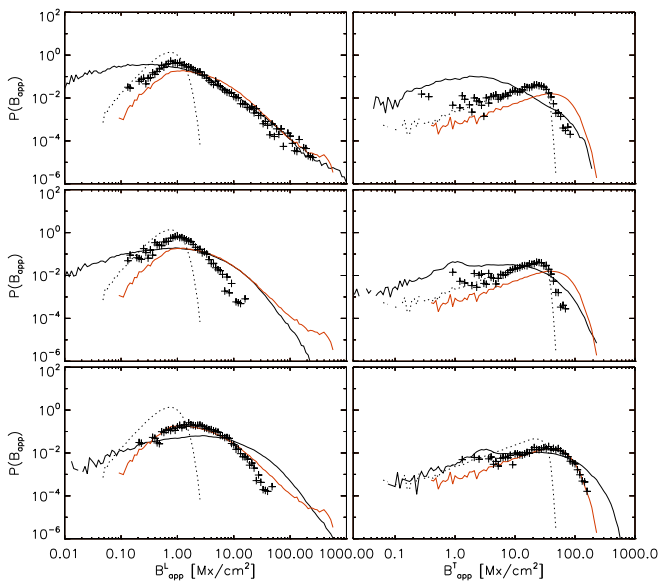}
    \caption{Probability density function (PDFs) for the longitudinal (left column) and transversal (right column) apparent magnetic flux density.
    PDFs from synthetic $B_{app}$ at original (solid)
    and Hinode (plus signs) resolution are compared with the observed PDFs from data set II (red solid line).
    \textit{From top to bottom}: mixed polarity simulation ($20$~G), dynamo run C and the same snapshot
    scaled by a factor 3. Also shown are the PDFs derived from Stokes parameters resulting from pure white noise with a standard deviation of $3\times10^{-4}$ (dotted lines).}
    \label{fig:pdfs}
\end{figure}

\citet{Graham:etal:2009} demonstrated how different effects can
influence the probability density function (PDF) of magnetic field
derived from Stokes V spectra, such that it differs from the PDF
of the underlying magnetic field. In particular, the effect of
noise leads to a PDF which has a peak at the position that
corresponds to the noise level. Here we compare the PDFs of
$B^{L}_{app}$ (left-hand panels) and $B_{app}^{T}$ (right-hand
panels) determined from the synthesized Stokes profiles with those
from observations. Figure~\ref{fig:pdfs} shows, from top to
bottom, PDFs computed for the mixed polarity ($20$~G) snapshot, a
snapshot from the dynamo run C without scaling and from the dynamo
run C after scaling by a factor of 3. Overplotted are the results
calculated from pure noise with a standard deviation corresponding
to the noise level of the observations (dotted lines) and the PDFs
obtained from data set II (red lines).

The figure shows that all the PDFs from the simulation snapshots
at Hinode resolution are strongly influenced by the noise at the
smallest field strengths, which is also the case for the PDF based
on the Hinode data. The noise-induced maxima lie at approximately
$B^{L}_{app}\approx 1$~G and $B_{app}^{T}\approx 20$~G. Due to the
mixed polarity field on the small scales, dynamo run C shows
significant loss of stronger signals after spatial smearing. The
mixed-polarity simulation snapshot, on the other hand, contains
larger unipolar patches so that it retains a considerable amount
of the stronger vertical field. Its PDF for $B_{app}^{L}$ has an
extended tail of stronger field, which corresponds to the observed
distribution. In the observations this extended tail is a result
of the contribution of the network flux concentrations. The
$B_{app}^{T}$ distributions of both, mixed-polarity and dynamo C
run snapshots, follow closely the distribution generated from the
pure noise, which means that the signature of the horizontal field
is mostly lost in the noise. Only a small percentage of the pixels
show a $B^{T}_{app}$ signal above the noise level. However, after
scaling the original field values by a factor 3, the PDF for
$B_{app}^{T}$ (bottom row of the Fig.~\ref{fig:pdfs}) agrees well
with the observed one. Small discrepancies at the smallest signals
result from poor sampling (a consequence of the small region
covered by the simulations). The corresponding synthesized
distribution of $B_{app}^{L}$ follows the observed distribution up
to approximately $10$~Mx/cm$^{2}$. Pixels with $B_{app}^{L}$
signals higher then $40$~Mx/cm$^{2}$ are very rare for the maps
resulting from dynamo run C.

\subsection{Changes with the heliocentric angle}

The variation of the apparent magnetic flux density at different
heliocentric angles has been determined using a snapshot from the
dynamo run C. The magnetic field strength is scaled by a factor of
3 in keeping with the findings from the previous sections. The
upper panels of Fig.~\ref{fig:b_cl} show height profiles of the
mean transversal (right) and mean absolute longitudinal (left)
field in the simulation at different heliocentric angles. Since
the transversal component of the magnetic field becomes
increasingly dominant in the layers above optical depth unity, the
mean absolute longitudinal (line-of-sight) component of the field
increases as the line of sight becomes inclined with respect to
the surface normal. However, the polarization signals in spectral
lines reflect the component of magnetic field averaged over the
line formation heights. The lower left panel of
Fig.~\ref{fig:b_cl} shows the PDF of $B_{ave}$, the vertical
component of magnetic field averaged over the height range that
corresponds to $\log \tau = [-3.5,0.1]$ at different heliocentric
angles. The PDFs at $\mu<0.8$ follow closely the PDF at $\mu=1$,
with a discrepancy at the strong field end. This discrepancy
increases with the heliocentric angle.

As illustrated by \citet{Graham:etal:2009}, $B_{ave}$ and
$|B_{app}^{L}|$ are well correlated, which explains the similar
trend of $|B_{app}^{L}|$ with the heliocentric angle, as shown in
the lower right panel of Fig.~\ref{fig:b_cl}. Plotted are the mean
values of $|B_{app}^{L}|$ and $B_{app}^{T}$ as functions of the
heliocentric angle. The mean value of $|B_{app}^{L}|$ stays
roughly constant until $\mu=0.8$ and then gradually decreases when
moving further away from disk center. The mean value of
$B_{app}^{T}$ decreases monotonically.

\subsection{Comparison with observations at $\mu=0.4$ ($\theta=66^{\circ}$)}

\begin{figure*}
    \includegraphics[width=0.34\hsize,angle=90]{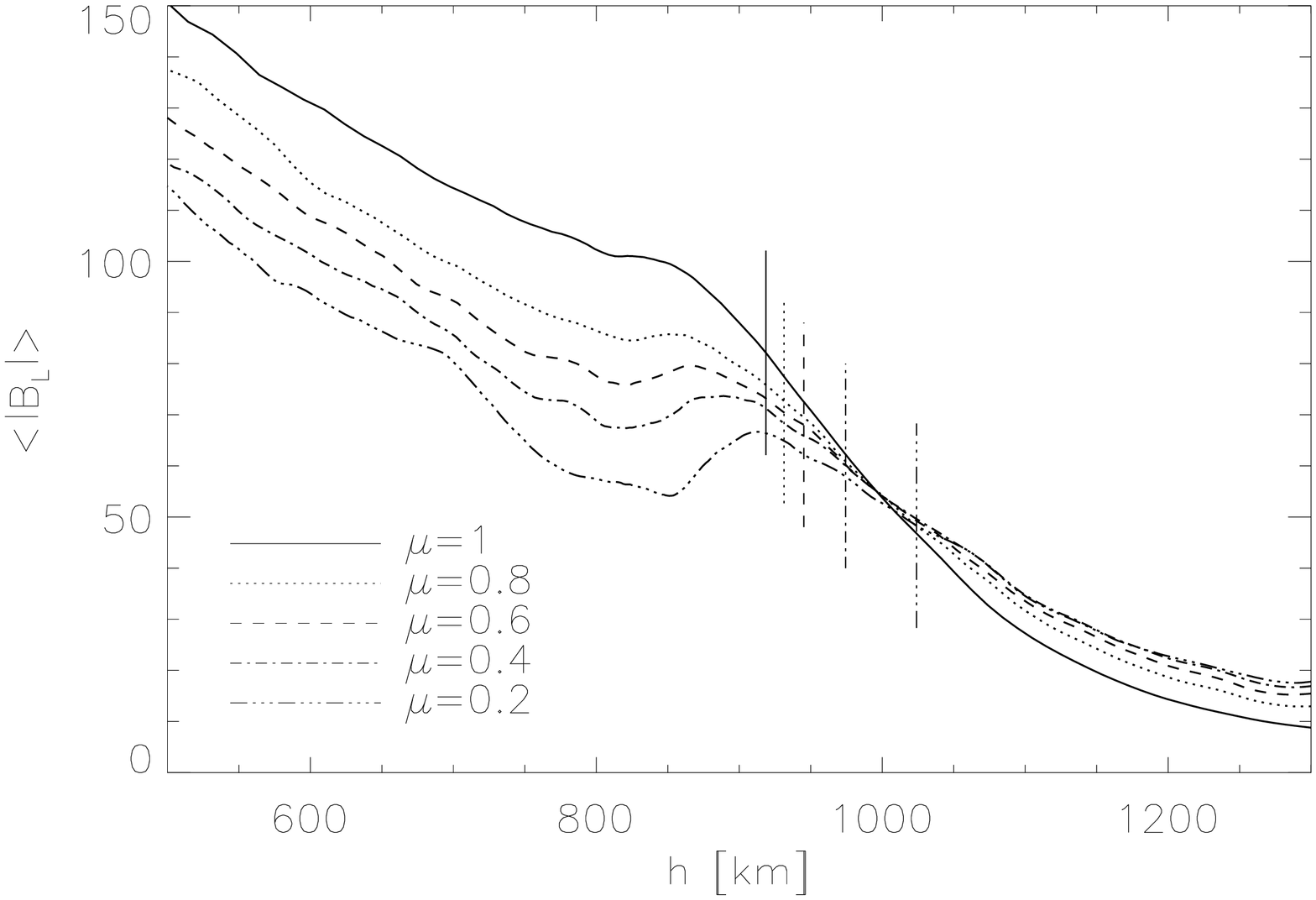}
    \includegraphics[width=0.34\hsize,angle=90]{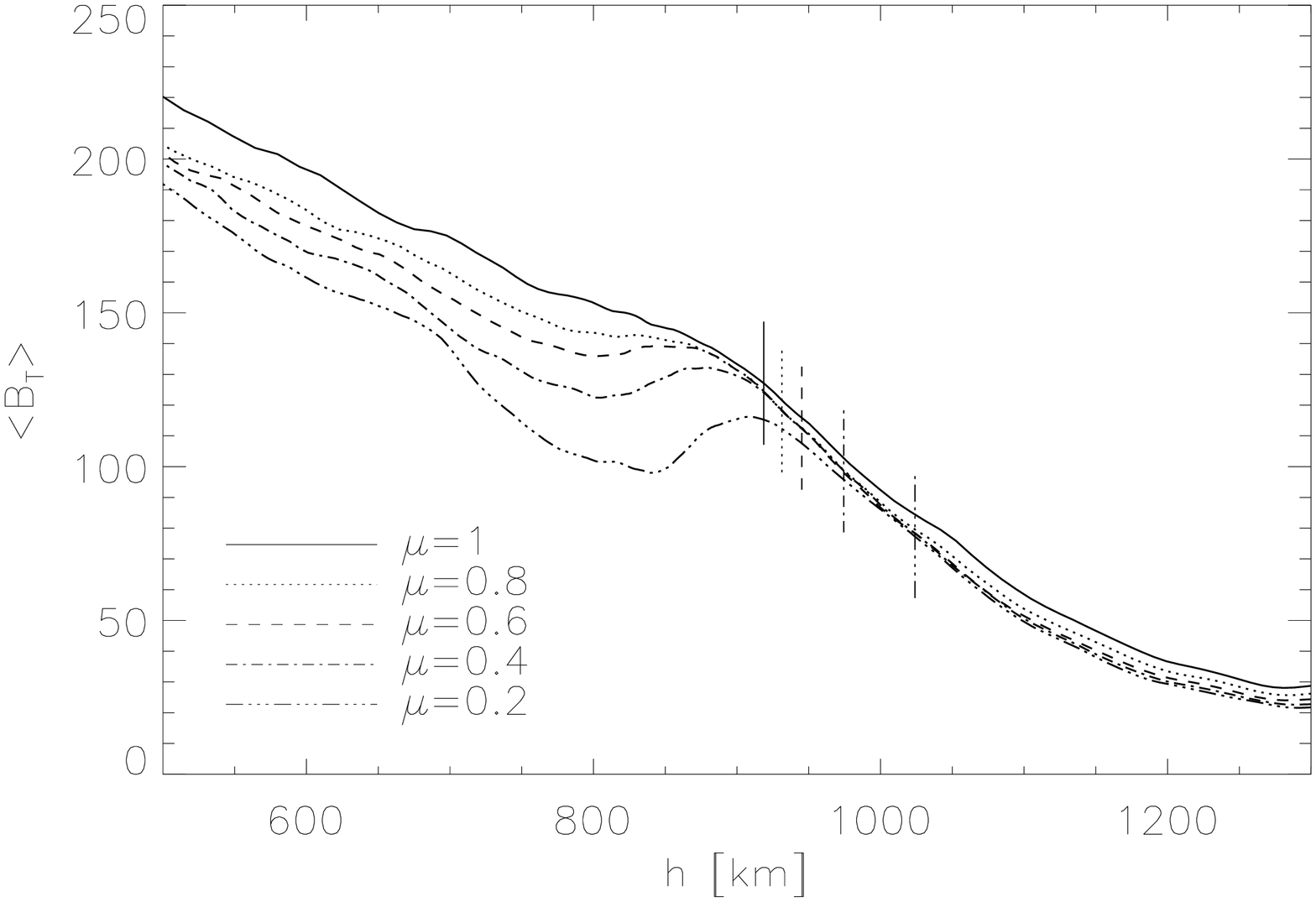}
    \includegraphics[width=0.34\hsize,angle=90]{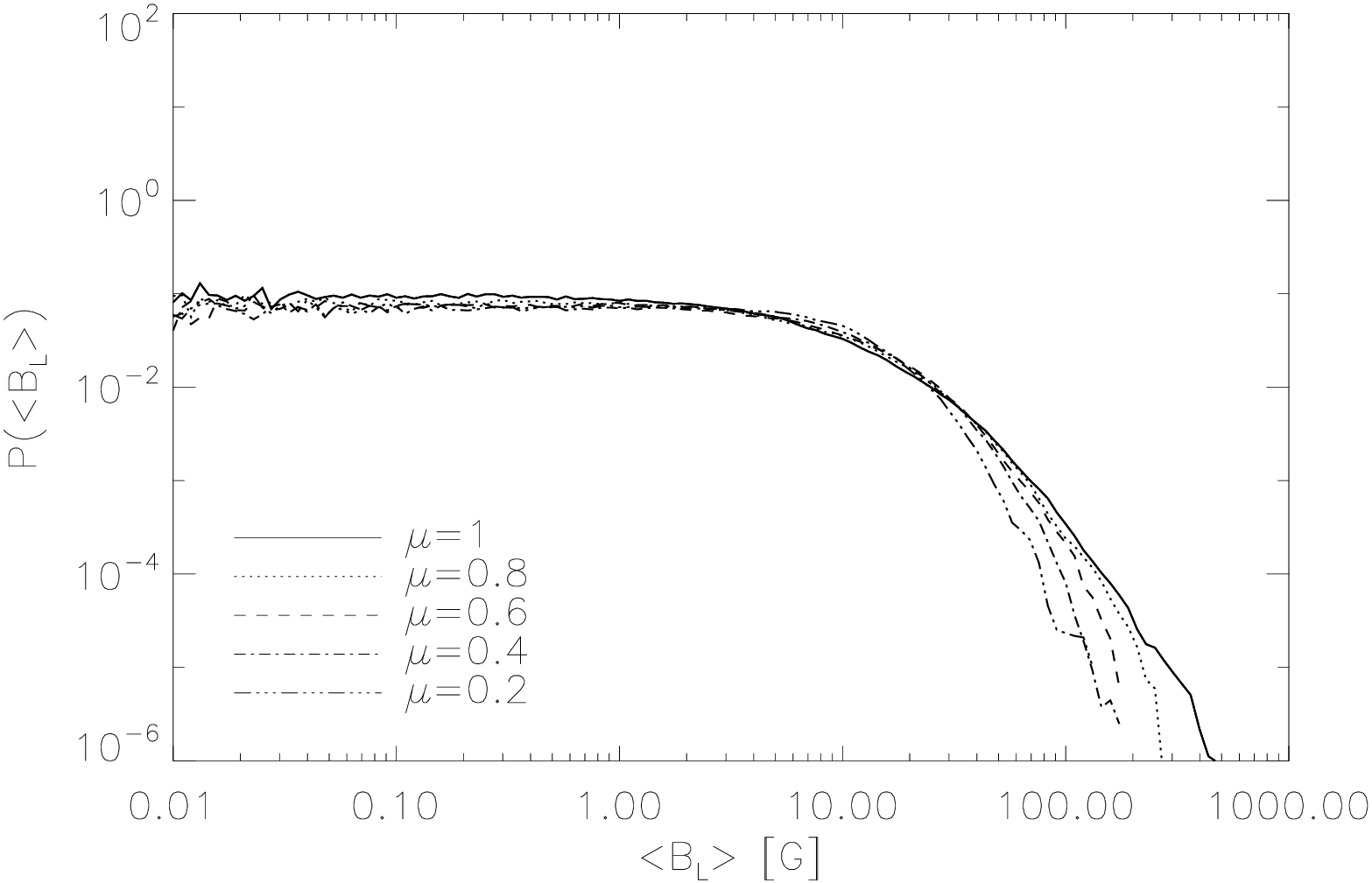}
    \includegraphics[width=0.36\hsize,angle=90]{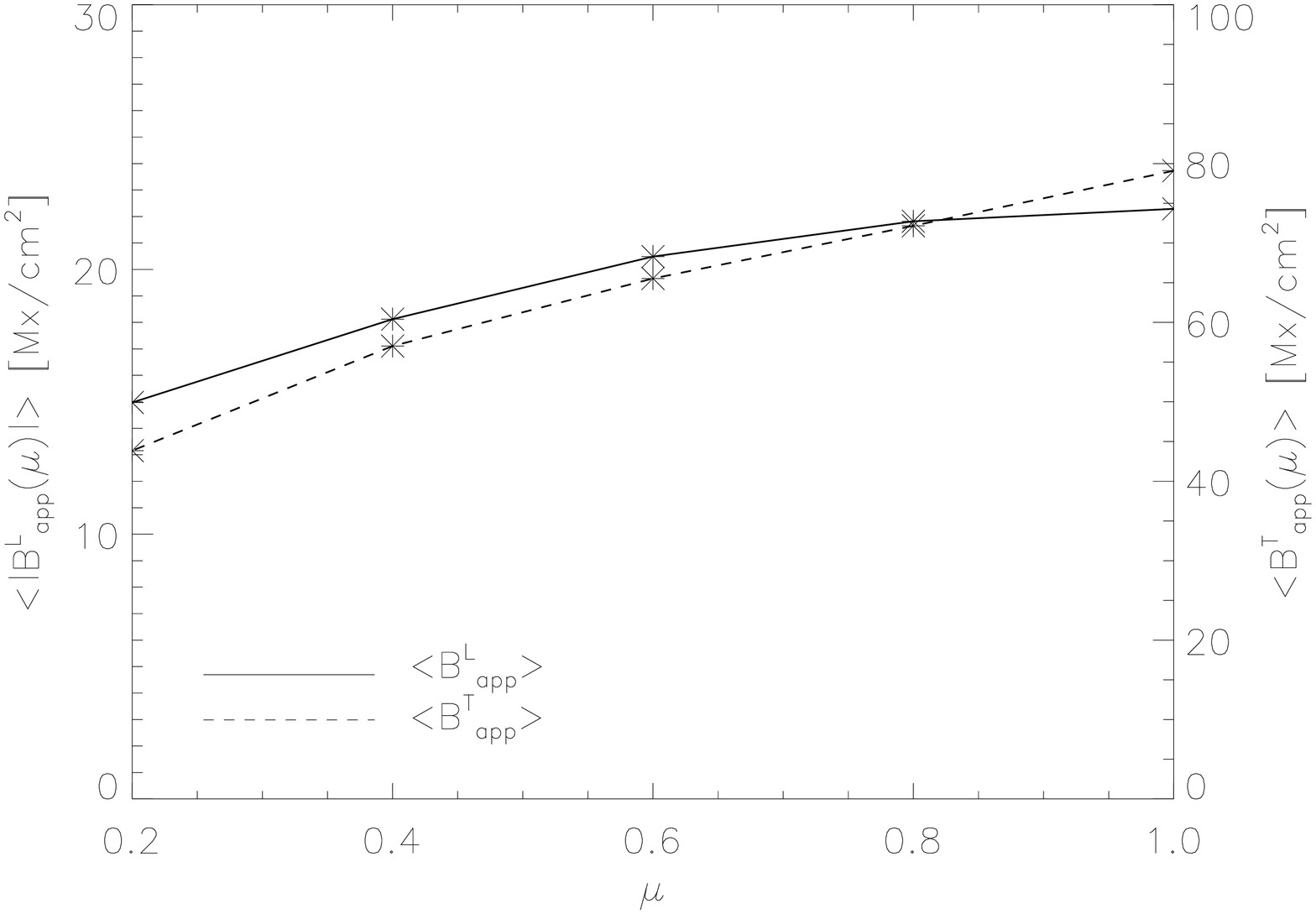}
    \caption{Simulated change with heliocentric angle retrieved from a dynamo run C snapshot.
    \textit{Upper panels:} The mean absolute longitudinal (left) and transversal (right) components of the magnetic field as a function of height in the simulation box at different heliocentric
    angles. Vertical lines indicate the level of optical depth unity.
    \textit{Lower left:} PDFs for the signed longitudinal field (averaged over the line formation height) at different heliocentric
    angles.
    \textit{Lower right:} Change of the mean absolute longitudinal ($|B_{app}^{L}|$) and mean transversal ($B_{app}^{T}$) apparent flux density as a function of the heliocentric angle.
    These values are calculated from synthesized Stokes profiles.}
    \label{fig:b_cl}
\end{figure*}

\begin{figure}
    \centering
    \includegraphics[angle=90,width=0.99\hsize]{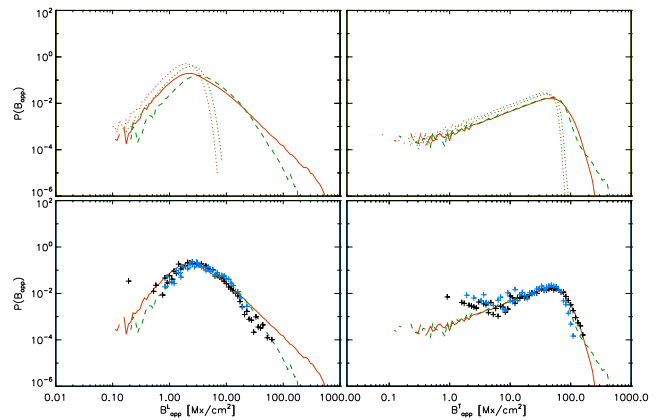}
    \caption{PDFs for the longitudinal (left column) and transversal (right column) apparent magnetic flux density. \textit{Upper row}: $B_{app}$ distributions obtained from observations
    at heliocentric angles $\theta=0^{\circ}$ (red solid line) and $\theta=66^{\circ}$ (green dashed line).
    The dotted lines in the same colors give the PDFs resulting from the corresponding noise.
    \textit{Bottom row}: Comparison of observations with the synthetic
    $B_{app}$ from dynamo run C (scaled with a factor 3) at heliocentric angles $\theta=0^{\circ}$ (black plus signs) and $\theta=66^{\circ}$ (blue
    plus signs), at Hinode resolution.}
    \label{fig:pdfs_cl}
\end{figure}

In this section, we compare the synthesized signals at Hinode
resolution with the observations. The upper panel of
Fig.~\ref{fig:pdfs_cl} shows PDFs of the field proxies,
$B^{L}_{app}$ and $B^{T}_{app}$, retrieved from the observations
at disc center and $\theta=66^{\circ}$ (data sets I and III),
respectively. The maxima of the PDFs are displaced with respect to
each other, which is due to the difference in the noise level
(Table~\ref{tab:obs}) as indicated by the overplotted PDFs
resulting from the pure noise. The probability of occurrence of
the $B_{app}^{L}$ signals larger than $30$~Mx/cm$^{2}$ decreases
at $\theta=66^{\circ}$, while the probability of occurrence of the
$B_{app}^{T}$ signals larger than $200$~Mx/cm$^{2}$ increases.
This effect results from the strong magnetic concentrations
present in the field of view. When a strong magnetic tube-like
structure with a vertical field of kG strength is observed under
an angle then $B_{app}^{L}$ signal decreases, while $B_{app}^{T}$
signal displays a corresponding increase relative to the value at
the disc center. This is because, projected on to the line of
sight, the strong vertical component of magnetic field gives rise
to a significant transversal component, leading to a strong
$B_{app}^{T}$ signal. Note, however that more pixels has values in
the range $70<B_{app}^{T}<200$~Mx/cm$^{2}$ where
PDF($B_{app}^{T}$) decreases towards the limb. Hence the mean
value of $B_{app}^{T}$ decreases with heliocentric angle, in
agreement with the result obtained from local dynamo simulations
at original resolution (shown in lower right frame of
Fig.~\ref{fig:b_cl}).

The lower panels of Fig.~\ref{fig:pdfs_cl} show the PDFs retrieved
from the dynamo run C snapshot at Hinode resolution. Their trends
correspond to the ones at original resolution, showing a small
difference only for the strong signals. The PDF of the synthesized
$B_{app}^{T}$ signal follows the observations at both heliocentric
angles up to $B_{app}^{T}\approx200$~Mx/cm$^{2}$. The synthesized
$B_{app}^{L}$ PDF matches the observations at $\theta=66^{\circ}$
up to approximately $20$~Mx/cm$^{2}$.

\section{Discussion}

We have based our comparison between MHD simulations and
observational results upon the magnetic proxies, the longitudinal
and transversal apparent magnetic flux density introduced by
\citet{Lites:etal:2008}. It has been shown that these proxies are
prone to the influence of the non-magnetic properties of the
atmosphere \citep{Beck:Rezaei:2009} and do not reflect correctly
the properties of the underlying field \citep{Graham:etal:2009}.
However, if we take that realistic MHD simulations represent the
solar atmosphere properly (supported by numerous studies, e.g.,
\citet{Schuessler:etal:2003,Shelyag:etal:2004,Shelyag:etal:2007,Danilovic:etal:2009})
and that our forward modelling of the instrumental effects and the
noise is correct, we introduce the same assumption as in the case
of the observations, allowing us to make a relatively unbiased
comparison with the findings of \citet{Lites:etal:2008}.

The magnetic proxies obtained from the synthesized data are
compared with observations of the internetwork region. The mixed
polarity simulations (which have a magnetic Reynolds number below
the threshold for dynamo action) reproduce the observed vertical
flux density, as was previously shown by
\citet{Khomenko:etal:2005a}. However, they do not contain enough
horizontal field, to be consistent with the Hinode data.
Simulations of surface dynamo give a ratio of the horizontal to
the vertical flux density consistent with the observational
results \citep{Schuessler:Voegler:2008}, but the overall
amplitudes are too low, at least for the simulations with the
magnetic Reynolds number feasible at the moment. A simple scaling
of the field in the simulation domain by a fixed factor brings the
magnetic flux density from the dynamo snapshots at Hinode
resolution close to the observed ones. Some justification for this
tentative procedure comes from comparing simulations with
different magnetic Reynolds numbers. The resulting mean magnetic
field strength is of the order of $70$~G at $\tau=0.1$ and around
$170$~G at the solar surface. This is roughly consistent with the
estimates given by \citet{Trujillo:etal:2004}, since the line they
consider samples the middle photosphere, and with extrapolations
based on the cancellation coefficient \citep{Graham:etal:2009}.

The contribution of the surface dynamo dominates in the regions
with $|B_{app}^{L}|<30$~Mx/cm$^{2}$ and
$B_{app}^{T}<200$~Mx/cm$^{2}$ at Hinode spatial resolution.
Everything stronger than that could imply: (1) flux emergence with
kG horizontal fields \citep{Cheung:etal:2008} or (2) strong
network fields, which have a different source from those in the
internetwork, a possibility supported by studies of ephemeral
active regions \citep{Harvey:1993,Hagenaar:etal:1999}.

The synthesized magnetic proxies exhibit a decrease from the disk
center to the limb. This is in qualitative agreement with the
observations by \citet{Lites:etal:2008}. The PDFs show little
$\mu$ dependence for the small values of the magnetic proxies, in
accordance with the results of \citet{Marian:etal:2008}. In
contrast, \citet{Steiner:etal:2009} could not reproduce the
observed simultaneous decrease towards the limb of the
longitudinal and transversal field proxies on the basis of MHD
simulations assuming a prescribed vertical magnetic flux or a
transport of horizontal flux into the computational box from
below.

Our results support the presence of local dynamo action in the
quiet Sun. They also suggest that the magnetic field is dominantly
horizontal. The ratio of the mean horizontal and vertical
component of the dynamo-generated magnetic field reaches values
between 2 and 4 in the optical depth interval $-2<\log \tau< -1$.
This gives the observed ratio of the transversal and longitudinal
apparent magnetic flux densities at the Hinode resolution when an
adequate noise level is considered.

We note, however, that the conclusions drawn here are based on a
simple assumption drawn from the properties of magnetic field in
the simulations with different magnetic Reynolds numbers. The
increase in the field strength by a factor of 2 or 3 would result
in a change of physical parameters as well as the dynamics of the
plasma since the strong field is much less susceptible to the
influence of the turbulent motions. These effects are not taken
into account in our scaling procedure. Thus, we consider our work
to be a step towards revealing the role of the surface dynamo in
the quiet Sun magnetism until more realistic simulations, as well
as higher resolution observation of the solar photosphere become
available.

\begin{acknowledgements}
We thank Jonathan Pietarila Graham for the valuable help with the
PDFs and Anna Pietarila for a critical reading of the paper.
Hinode is a Japanese mission developed and launched by ISAS/JAXA,
with NAOJ as domestic partner and NASA and STFC (UK) as
international partners. It is operated by these agencies in
co-operation with ESA and NSC (Norway). This work was partly
supported by WCU grant No. R31-100016 from Korean Ministry of
Education, Science and Technology.
\end{acknowledgements}

\bibliographystyle{aa}

\begin{thebibliography}{}

\bibitem[Asensio Ramos(2009)]{Asensio:2009} Asensio Ramos, A.\ 2009,
\apj, 701, 1032
\bibitem[Asensio Ramos et al.(2007)]{Asensio:etal:2007} Asensio Ramos,
A., Mart{\'{\i}}nez Gonz{\'a}lez, M.~J., L{\'o}pez Ariste, A. et
al. \apj, 659, 829
\bibitem[Beck \& Rezaei(2009)]{Beck:Rezaei:2009} Beck, C., \& Rezaei, R.\ 2009,
\aap, 502, 969
\bibitem[Bello Gonz{\'a}lez et al.(2009)]{Nazaret:etal:2009} Bello Gonz{\'a}lez, N., Yelles
Chaouche, L., Okunev, O., \& Kneer, F.\ 2009, \aap, 494, 1091
\bibitem[Berdyugina \& Fluri(2004)]{Berdyugina:Fluri:2004} Berdyugina, S.~V., \&
Fluri, D.~M.\ 2004, \aap, 417, 775
\bibitem[Cattaneo(1999)]{Cattaneo:1999} Cattaneo, F.\ 1999, \apjl,
515, L39
\bibitem[Centeno et al.(2007)]{Centeno:etal:2007} Centeno, R., et al.\
2007, \apjl, 666, L137
\bibitem[Cheung et al.(2008)]{Cheung:etal:2008} Cheung, M.~C.~M.,
Sch{\"u}ssler, M., Tarbell, T.~D., \& Title, A.~M.\ 2008, \apj,
687, 1373
\bibitem[Danilovic et al.(2008)]{Danilovic:etal:2008} Danilovic, S., Gandorfer, A.,
Lagg, A., et al. \ 2008, \aap, 484, L17
\bibitem[Danilovic et al.(2009)]{Danilovic:etal:2009} Danilovic, S.,
Sch{\"u}ssler, M., \& Solanki, S.~K.\ 2009, arXiv:0910.1211
\bibitem[Dom{\'{\i}}nguez Cerde{\~n}a et
al.(2003)]{DC:etal:2003} Dom{\'{\i}}nguez Cerde{\~n}a, I.,
S{\'a}nchez Almeida, J., \& Kneer, F.\ 2003, \aap, 407, 741
\bibitem[Frutiger et al.(2000)]{Frutiger:etal:2000} Frutiger, C., Solanki, S.~K.,
Fligge, M., \& Bruls, J.~H.~M.~J.\ 2000, \aap, 358, 1109
\bibitem[Hagenaar et al.(1999)]{Hagenaar:etal:1999} Hagenaar, H.~J.,
Schrijver, C.~J., Title, A.~M., \& Shine, R.~A.\ 1999, \apj, 511,
932
\bibitem[Harvey(1993)]{Harvey:1993} Harvey, K.~L.\ 1993,
Ph.D.~Thesis, Utrecht University
\bibitem[Harvey et al.(2007)]{Harvey:etal:2007} Harvey, J.~W., Branston,
D., Henney, C.~J., \& Keller, C.~U.\ 2007, \apjl, 659, L177
\bibitem[Ishikawa et al.(2008)]{Ishikawa:etal:2008} Ishikawa, R., et al.\ 2008, \aap,
481, L25
\bibitem[Kosugi et al.(2007)]{Kosugi:etal:2007} Kosugi, T., et al.\
2007, \solphys, 243, 3
\bibitem[Khomenko et al.(2003)]{Khomenko:etal:2003} Khomenko, E.~V., Collados, M.,
Solanki, S.~K., et al. \ 2003, \aap, 408, 1115
\bibitem[Khomenko et al.(2005a)]{Khomenko:etal:2005a} Khomenko, E.~V., Mart{\'{\i}}nez
Gonz{\'a}lez, M.~J., Collados, M., et al. \ 2005, \aap, 436, L27
\bibitem[Khomenko et al.(2005b)]{Khomenko:etal:2005b} Khomenko, E.~V., Shelyag, S.,
Solanki, S.~K., V{\"o}gler, A.\ 2005, \aap, 442, 1059
\bibitem[Lites et al.(2001)]{Lites:etal:2001} Lites, B.~W., Elmore,
D.~F., \& Streander, K.~V.\ 2001, in: ASP conf. ser., Vol 236,
Advanced Solar Polarimetry -- Theory, Observation, and
Instrumentation, ed. M. Sigwarth, 33
\bibitem[Lites et al.(2008)]{Lites:etal:2008} Lites, B.~W., et al.\
2008, \apj, 672, 1237
\bibitem[Martin(1988)]{Martin:1988} Martin, S.~F.\ 1988, \solphys,
117, 243
\bibitem[Mart{\'{\i}}nez Gonz{\'a}lez et al.(2007)]{Marian:etal:2007} Mart{\'{\i}}nez Gonz{\'a}lez, M.~J.,
Collados, M., Ruiz Cobo, B., \& Solanki, S.~K.\ 2007, \aap, 469,
L39
\bibitem[Mart{\'{\i}}nez Gonz{\'a}lez et al.(2008)]{Marian:etal:2008} Mart{\'{\i}}nez Gonz{\'a}lez, M.~J.,
Asensio Ramos, A., L{\'o}pez Ariste, A., \& Manso Sainz, R.\ 2008,
\aap, 479, 229
\bibitem[Orozco Su{\'a}rez et
al.(2007a)]{Orozco:etal:2007a} Orozco Su{\'a}rez, D., et al.\
2007, \apjl, 670, L61
\bibitem[Orozco Su{\'a}rez et al.(2007b)]{Orozco:etal:2007b} Orozco
Su{\'a}rez, D., et al.\ 2007, \pasj, 59, 837
\bibitem[Orozco Su{\'a}rez et al.(2008)]{Orozco:etal:2008:vfig} Orozco Su{\'a}rez, D., Bellot
Rubio, L.~R., del Toro Iniesta, J.~C., \& Tsuneta, S.\ 2008, \aap,
481, L33
\bibitem[Pietarila Graham et al.(2009)]{Graham:etal:2009} Pietarila
Graham, J., Danilovic, S., \& Sch{\"u}ssler, M.\ 2009, \apj, 693,
1728
\bibitem[Petrovay \& Szakaly(1993)]{Petrovay:Szakaly} Petrovay, K., \& Szakaly,
G.\ 1993, \aap, 274, 543
\bibitem[S{\'a}nchez Almeida et al.(2003)]{SA:etal:2003} S{\'a}nchez
Almeida, J., Emonet, T., \& Cattaneo, F.\ 2003, \apj, 585, 536
\bibitem[Sch{\"u}ssler et al.(2003)]{Schuessler:etal:2003} Sch{\"u}ssler,
M., Shelyag, S., Berdyugina, S., et al. \ 2003, \apjl, 597, L173
\bibitem[Shelyag et al.(2007)]{Shelyag:etal:2007} Shelyag, S.,
Sch{\"u}ssler, M., Solanki, S.~K., V{\"o}gler, A.\ 2007, \aap,
469, 731
\bibitem[Shelyag et al.(2004)]{Shelyag:etal:2004}
Shelyag, S., Sch{\"u}ssler, M., Solanki, S.~K. et al., \ 2004,
\aap, 427, 335
\bibitem[Solanki(2009)]{Solanki:2009} Solanki,
S.~K.\ 2009, Astronomical Society of the Pacific Conference
Series, 405, 135
\bibitem[Steiner et al.(2008)]{Steiner:etal:2008} Steiner, O., Rezaei,
R., Schaffenberger, W., \& Wedemeyer-B{\"o}hm, S.\ 2008, \apjl,
680, L85
\bibitem[Steiner et al.(2009)]{Steiner:etal:2009} Steiner, O., Rezaei,
R., Schlichenmaier, R., et al. \ 2009, arXiv:0904.2030
\bibitem[Piskunov et al.(1995)]{Piskunov:etal:1995} Piskunov, N.~E., Kupka, F.,
Ryabchikova, T.~A., et al. \ 1995, \aaps, 112, 525
\bibitem[Thevenin(1989)]{Thevenin:1989} Thevenin, F.\ 1989, \aaps, 77, 137
\bibitem[Trujillo Bueno et al.(2004)]{Trujillo:etal:2004} Trujillo Bueno,
J., Shchukina, N., \& Asensio Ramos, A.\ 2004, \nat, 430, 326
\bibitem[Tsuneta et al.(2008)]{Tsuneta:etal:2008} Tsuneta, S., Ichimoto, K., Katsukawa, Y. et al.\ 2008, \solphys, 249, 167
\bibitem[V{\"o}gler et  al.(2005)]{Voegler:etal:2005} V{\"o}gler, A., Shelyag, S., Sch{\"u}ssler, M., et al. \ 2005, \aap, 429, 335
\bibitem[V{\"o}gler(2003)]{Voegler:2003} V{\"o}gler, A.\ 2003, PhD Thesis, University of G{\"o}ttingen, Germany,
http://webdoc.sub.gwdg.de/diss/2004/voegler
\bibitem[V{\"o}gler \& Sch{\"u}ssler(2007)]{Voegler:Schuessler:2007} V{\"o}gler, A.,
\& Sch{\"u}ssler, M.\ 2007, \aap, 465, L43
\bibitem[Sch{\"u}ssler \& V{\"o}gler(2008)]{Schuessler:Voegler:2008} Sch{\"u}ssler,
M., \& V{\"o}gler, A.\ 2008, \aap, 481, L5
\bibitem[de Wijn et al.(2008)]{deWijn:etal:2008} de Wijn, A.~G., Stenflo, J. O. Solanki, S. K. and Tsuneta, S. 2008, Space Sci. Rev. , Volume 144, Issue 1-4, pp. 275-315
\end{thebibliography}

\end{document}